\documentclass[fleqn]{ILASS_2012_Paper_Format}
\usepackage{xcolor}
\usepackage{graphicx}
\usepackage{subfigure}
\usepackage{floatflt}
\begin{document}
\title{Experiments and Direct Numerical Simulations \\
of binary collisions of miscible liquid droplets with different viscosities}
\runningheads{12$^{th}$ ICLASS 2012}{Experiments and Direct Numerical Simulations of binary collisions of miscible liquid droplets with different viscosities}
\author{C.~Focke$^{1}$, D.~Bothe$^{1}$\footnotemark[1], M.~Kuschel$^{2}$, M.~Sommerfeld$^{2}$ \footnotemark[2]}
\address{$^{1}$Center of Smart Interfaces, University of Darmstadt, Germany\\
focke@csi.tu-darmstadt.de and bothe@csi.tu-darmstadt.de\\
$^{2}$Mechanische Verfahrenstechnik, Martin-Luther-Universit\"at Halle-Wittenberg\\
matthias.kuschel@uni-halle.de and martin.sommerfeld@uni-halle.de\\
}
\footnotetext[1]{\normalsize{Corresponding author numerical part:
\href{mailto:bothe@csi.tu-darmstadt.de}{\textstyleInternetlink{bothe@csi.tu-darmstadt.de}}}}
\footnotetext[2]{\normalsize{Corresponding author experimental part:
\href{mailto:martin.sommerfeld@uni-halle.de}{\textstyleInternetlink{martin.sommerfeld@uni-halle.de}}}}
\abstract{
Binary droplet collisions are of importance in a variety of practical applications comprising dispersed two{-}phase flows. The background of our research is the prediction of properties of particulate products formed in spray processes. To gain a more thorough understanding of the elementary sub{-}processes inside a spray, experiments and direct numerical simulations of binary droplet collisions are used. The aim of these investigations is to develop semi{-}analytical descriptions for the outcome of droplet collisions. Such collision models can then be employed as closure terms for scale{-}reduced simulations.
In the present work we focus on the collision of droplets of different liquids. These kinds of collisions take place in every spray drying process when droplets with different solids contents collide in recirculation zones. A new experimental method has been developed allowing for high spatial and time resolved recordings via Laser{-}induced fluorescence. The results obtained with the proposed method will be compared with DNS simulations. The viscosities of the droplets are different whereas the interfacial tension and density are equal. The liquids are miscible and no surface tension is acting between the two liquids. Our intention is to discover elementary phenomena caused by the viscosity ratio of the droplets.} 
\section{Introduction}
In numerous industrial areas spray drying is used for producing powders of desired properties. The first step of this process is the atomization of a solution or suspension to produce fine droplets which are subsequently dried in a hot air stream. Normally, the atomisation and subsequent droplet break{-}up yield a broad size distribution of droplets. Owing to their size, the residence time of droplets will vary and their viscosity will be increased as a result of the drying process. During spray drying processes droplet collisions will occur not only close to the nozzle \cite{Shu06} but also in recirculation zones further downstream in the dryer. Here droplets of different drying state (i.e. different solids content and viscosity) may collide with each other. Depending on the relative velocity, the impact parameter and the diameter ratio, different collision outcomes can be observed \cite{Kro10} (i.e. bouncing, coalescence, separation). Due to the evaporation of the solvent, collisions of unlike viscous droplets will mostly exhibit also a diameter ratio less than unity. 
In the experiments a fluorescent marker is used to visualise mixing processes. This was already done for water or water ethanol mixtures in \cite{Ash90,Gao05}, but both studies had in common that they used a frozen image technique yielding only time averaged data. The observation of deformations, mixing and penetration processes of single droplets was not possible and hence some features belonging to a certain collision type were hidden. The proposed experimental method offers high time resolution and still a very good contrast where no in{-}motion unsharpness occurs. Thus it is possible to study not only the dynamical behaviour of the collision complex (outer surface) but also the internal mixing and even penetration can be analysed quantitatively. Another advantage which comes with the new method is based on the fact that numerical researchers now have better validation data due to the high time resolution.
In the numerical part of the work we investigate the collision of droplets of different liquids and assume that surface tension and density are equal. The liquids are miscible, so no surface tension is acting at the liquid{-}liquid interface. The experimental data is used to validate the effect of non{-}constant viscosity transported using a species equation. Subsequently, the local field data is used to gain a deep insight in the flow inside the colliding droplets. This study is devoted to the investigation of such collisions experimentally as well as numerically in order to analyse the effects of penetration and mixing.
\section{Experimental Facility}
\begin{figure}[hbtp]
\centering
\includegraphics[height=0.4\textwidth]{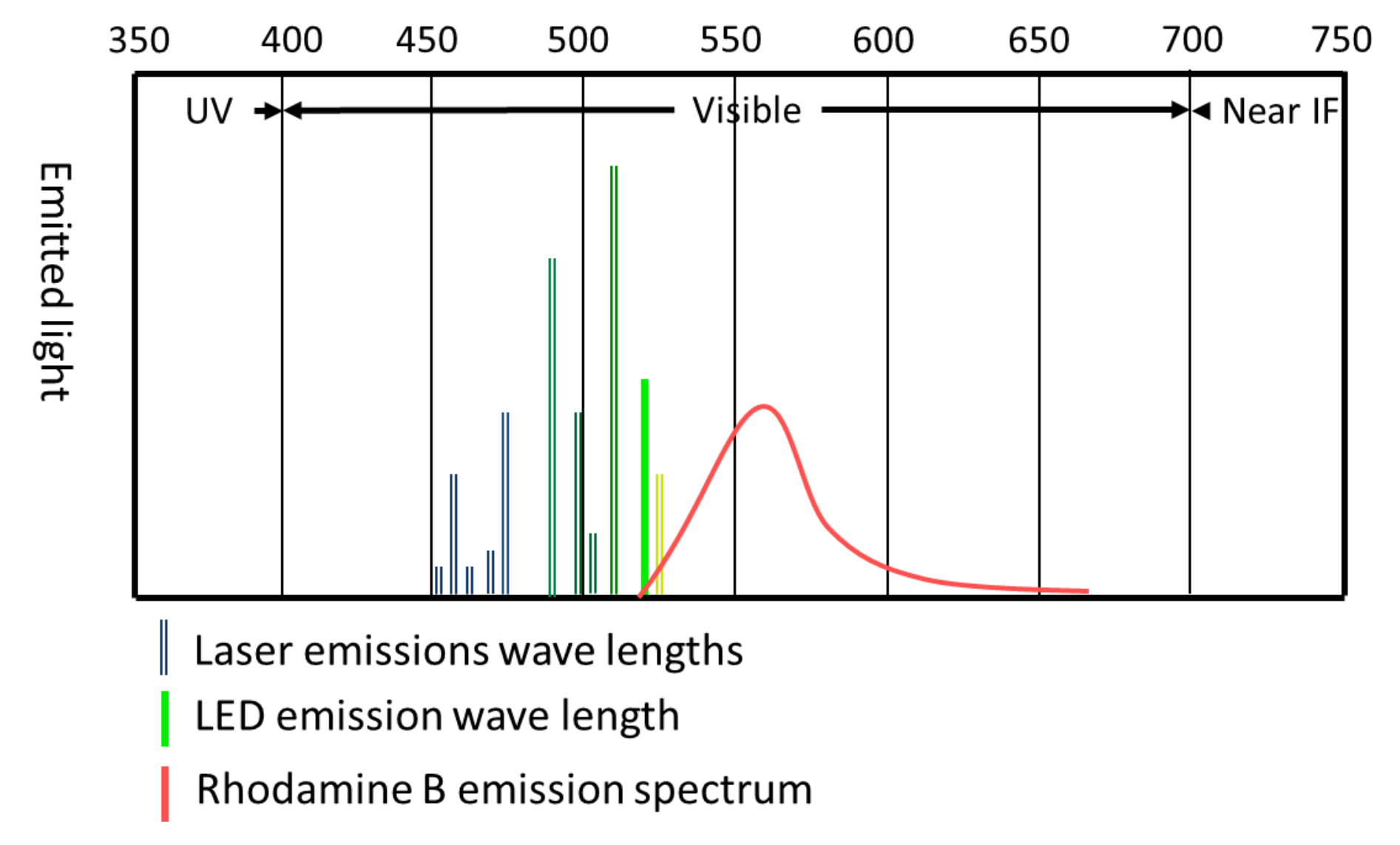}
\caption{Emissionsspectra of different light sources}
\label{fig:Emissionsspectra}
\end{figure}

\begin{figure}[hbtp]
\centering
\subfigure[Front view]
{\includegraphics[width=0.4\textwidth]{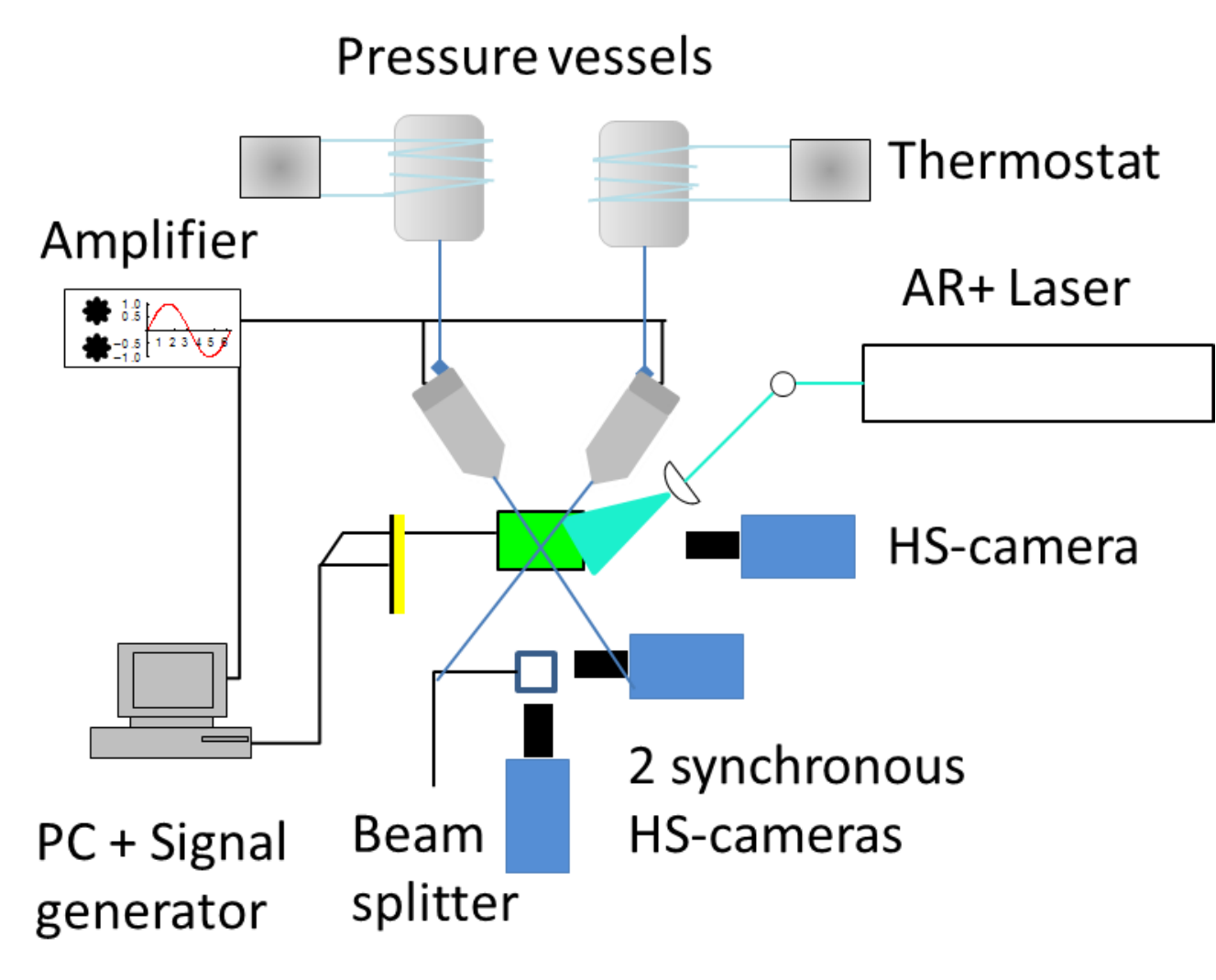}}
\subfigure[Top view]
{\includegraphics[width=0.4\textwidth]{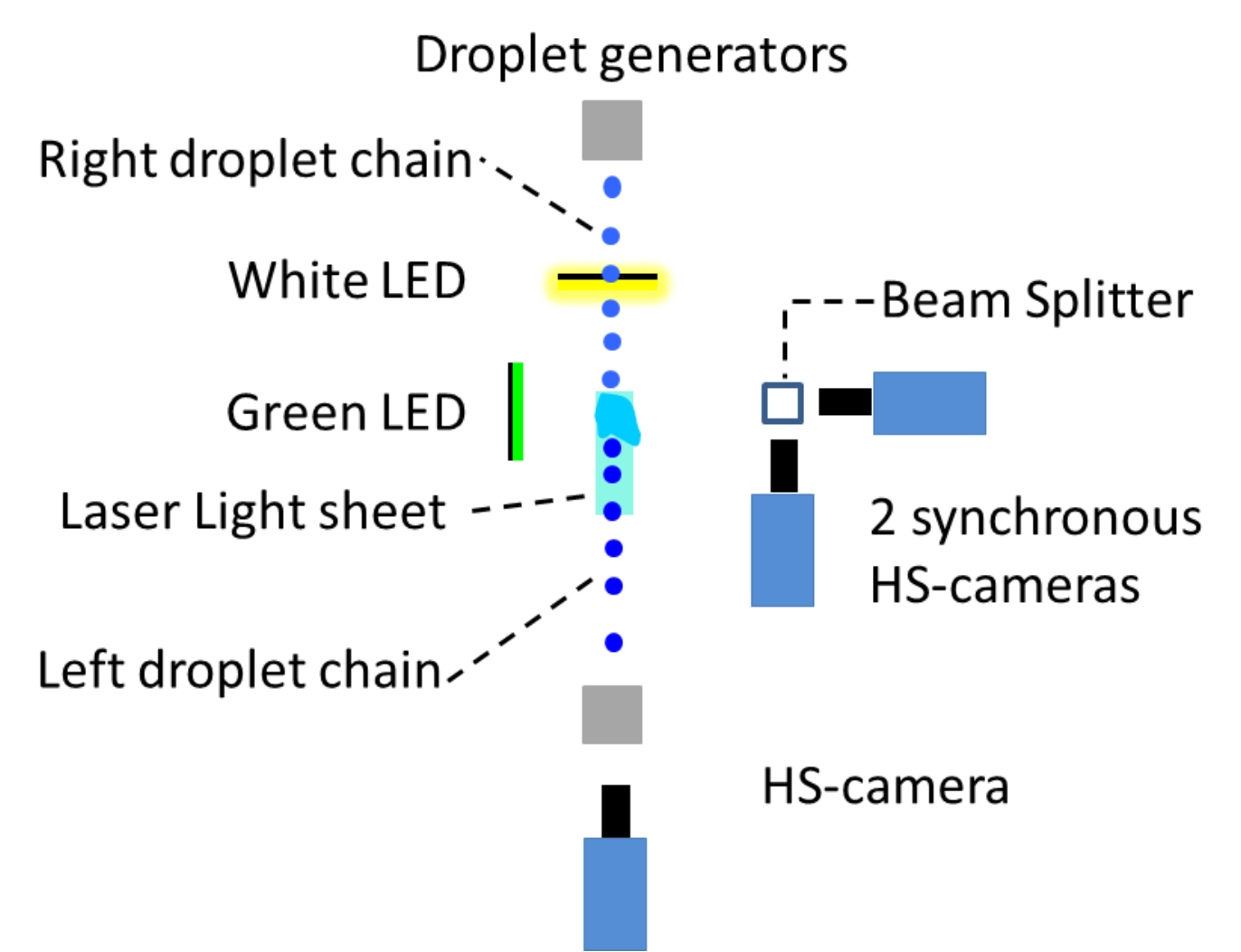}}
\caption{Experimental setup from different viewpoints}
\label{fig:Experimental setup}
\end{figure}

For the investigation of droplet collisions with equal und unequal viscosity, uniform droplets have to be produced which was realised by using two piezo{-}electric droplet generators. The excited jets break up into monodisperse droplets chains. The angle between the resulting droplet chains was varied in order to change the relative velocity in a range of 1 to 3 m/s. The liquid was pressed through the nozzles (producer: encap biosystems) with a diameter of 200 or 300 $\mu m$, resulting in droplets of around 380 and 580 $\mu m$ in size. The temperature was kept constant at about 22 $^\circ$C by a thermostat. The impact parameter was modified by using the aliasing method (frequency shift) \cite{Aliasinggotaas}. Three high speed cameras (two PHOTRON SA 4 and one PCO 1200 HS), whereas the two PHOTRON cameras operated synchronously, were used to observe the collision event from different viewpoints (two cameras front or collision plane view and one camera side view). The synchronous cameras were equipped with the same lens yielding a calibration factor of 11.86 $\mu m/Pixel$. Both cameras observed the collision from the same view point via a beam splitter (50/50) but with different requirements. Whereas one camera recorded a combination of green LED back light and fluorescence of the droplets, the other camera looked for the fluorescence of the single droplets and their mixture only. Therefore special glass filters (SCHOTT) were applied to meet the requirements (see Fig. \ref{fig:Emissionsspectra}). 
First, all filters had to extinguish the scattered laser light but had to be able to pass the LED back light. Additionally, the filter of the fluorescent recording camera also had to eliminate the LED back light. So the filters were chosen as follows: 530 $nm$ for the camera which recorded the combination of back light and fluorescence and a 590 $nm$ filter was equipped on the other camera. The PCO camera has been positioned parallel to the collision plane to assure central collisions in that plane and was also equipped with a 530 $nm$ filter to reduce the scattered laser light in order to protect the camera from too much light intensity. The mixing and penetration processes have been visualised with the help of the fluorescent marker Rhodamin B with a concentration of 200 $mg/kg_{Liquid}$. The Laser light sheet was created by a AR+ Laser (LEXEL 3500) and has been expanded to 20 mm in height and 1 mm in width at the collision point, generating a mean Laser intensity of around 150 $kW/m^2$ (see Fig.\ref{fig:Experimental setup}).
With the experimental setup it is now possible to observe the collision as well as the mixing/penetration process time{-} and spatial resolved due to the application of two synchronous operating cameras, which either record a combination fluorescent droplets and non{-}fluorescent droplets or only the fluorescence of the collision complex at arbitrary frame rates up to 30.000 frames per second and a shutter rate of 10 $\mu$s. 

\section{Numerical Methods}
\subsection{Model description}
The numerical method (FS3D) employed here is based on the Volume{-}of{-}Fluid (VOF) method by Hirt and Nichols~\cite{Hirt1981201}. The VOF method solves the Navier-Stokes equations for an incompressible transient two{-}phase flow. The fundamental idea of this method is to capture the interface position implicitly by means of a phase indicator function, i.e. a scalar function $f = f(t, \textbf{x})$ with, say, $f = 1$ in the dispersed phase and $f = 0$ in the continuous phase. Due to the absence of phase change, the transport of \textit{f} is governed by the advection equation
\begin{equation}
\partial_t f+\mathbf{u} \cdot \nabla f=0 .
\end{equation} 
Therefore, the VOF{-}method inherently conserves phase volume. This is an important issue especially if the long term behaviour of solutions is to be studied, where other methods like Level Set or Front tracking could lose too much of the droplet volume. In case of Level Set methods it is possible to include correction steps in order to conserve the phase volume~\cite{Olsson2005225}. In the Finite Volume (FV) discretisation scheme employed here, the cell centered value of \textit{f} corresponds to the phase fraction inside a computational cell. Based on these values, an approximation of the interface normal can be computed as
\begin{equation}
\mathbf{n}=\frac{\nabla f} {\| \nabla f \|}.
\label{eq:normals}
\end{equation} 
Combination of the fractional volume of the dispersed phase with the interface normal then allows for an interface reconstruction with the so{-}called PLIC method~\cite{Rider1998112}. This reconstruction of the phase geometry inside the computational cells creates important subgrid{-}scale information. The reconstructed interface is employed for an accurate convective transport of the phase indicator. Once the phase distribution is known, a one{-}field formulation of the Navier{-}Stokes system is possible in which the interfacial momentum jump condition acts as a source term in the momentum equations. It reads as
\begin{equation}
\rho \partial_t \mathbf{v} + \rho (\mathbf{v} \cdot \nabla)\mathbf{v} = -\nabla p + \nabla \cdot \mathbf{S} + \rho \mathbf{g} + \sigma \kappa \mathbf{n} \delta_\Sigma, 
\end{equation}
with the viscous stress tensor
\begin{equation}
\mathbf{S}=\eta \left( \nabla \mathbf{v} + \left( \nabla \mathbf{v} \right) ^{T} \right), 
\end{equation} 
where the material properties $\rho$ and $\eta$ refer to the phase dependent values. The local values of the density are determined by
\begin{equation}
\rho =\rho_d f + \rho_c (1-f) .
\end{equation} 
$\mathrm{\kappa}$ denotes the sum of the principal curvatures of the interface. It is given by
\begin{equation}
\kappa =-\nabla \cdot \mathbf{n}.
\end{equation} 
The interfacial Delta distribution $\mathrm{\delta_\Sigma}$ can be computed as
\begin{equation}
\delta_\Sigma=||\nabla f ||.
\end{equation} 
Hence, $\mathbf{n}_\Sigma \delta_\Sigma$ can be approximated as $\nabla f$, where $f$ is taken from a smoothed version of the phase indicator function field. For more details see~\cite{Foc12}.
\subsection{Extension of FS3D for the computation of non{-}isoviscous liquids}
\begin{figure}[!htb]
\begin{center} 
\includegraphics[width=8cm]{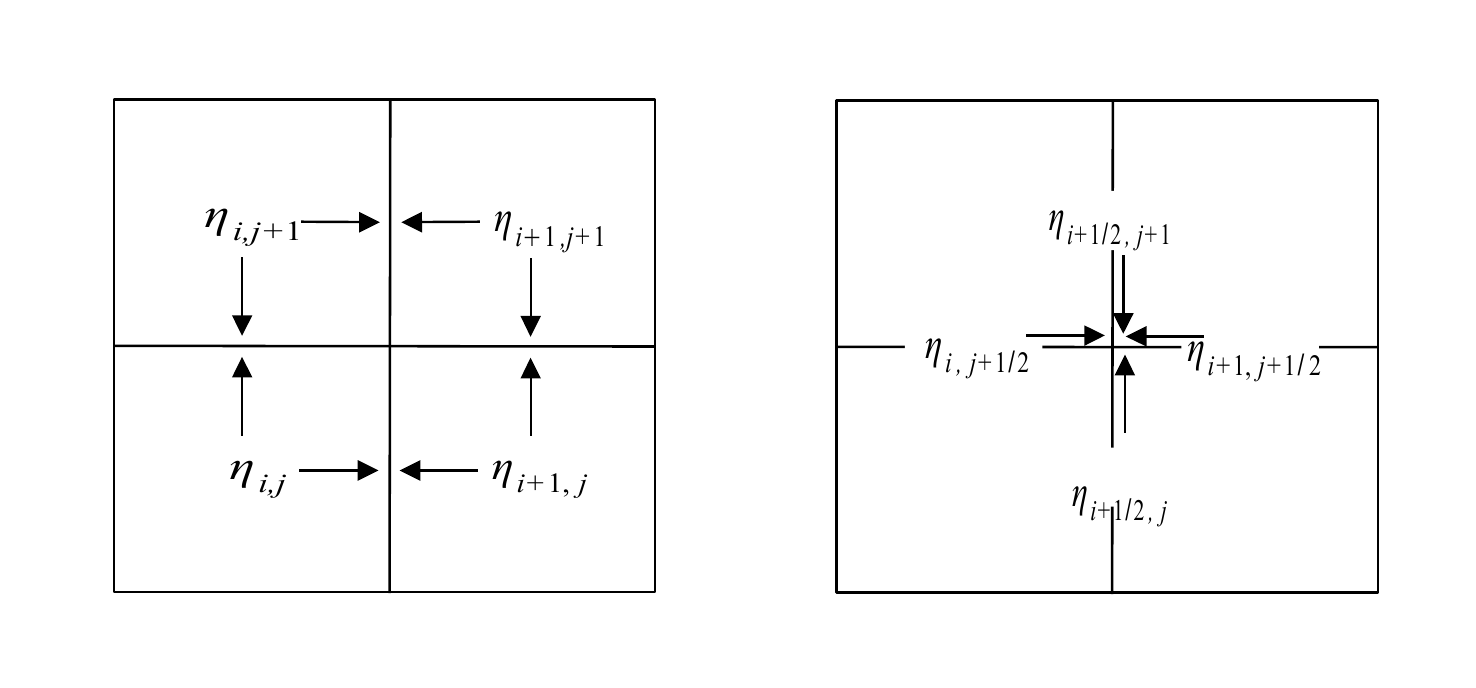}
\end{center}
\caption{Scheme of the viscosity computation at the liquid{-}liquid and the liquid{-}air interface.}
\label{fig:ViscosityAveraging}
\end{figure}

FS3D has been extended to simulate the collision of liquid droplets with different viscosities. The viscosity is modeled using a mixture viscosity model based on the local mass fraction $\mathrm{\varphi= \rho_1/\rho}$. The mass fraction is transported according
\begin{equation} 
\partial_{t} \varphi + \mathbf{v} \cdot \nabla \varphi + \nabla \cdot j_1 = 0,\quad j_1 = -D_1 \nabla \varphi ,
\label{eq:massfractiontransport}
\end{equation}
with the diffusion coefficient $\mathrm{D_1}$ of the liquid solvent in the liquid solute. 
The viscosity at the liquid{-}liquid interface is computed analogously to the treatment at the liquid{-}air interface, using an appropriate combination of arithmetic and harmonic means of the viscosity of both adjoining liquids. The consideration of the viscosity at the liquid{-}liquid and the liquid{-}air interface is different for different components of the stress. In case of extensional flow, central differences of the velocity on the adjacent cell faces are evaluated. For the calculation of the normal stress, a viscosity is required in the cell center. The viscosity in the cell center is computed according to
\begin{equation}
\eta = \eta_C \left( 1-f\right) + f \left( \eta_D + \left( \eta_\varphi -\eta_D \right) \right) \varphi ,
\label{eq:ViscosityAveraging}
\end{equation} 
where the index C corresponds to the continuous phase, D to the disperse phase and $\varphi$ to the mass fraction.
In case of friction due to shear flow, the shear rates are calculated as central differences of velocity. In contrast to the elongational flow, the velocities of two adjoining cell faces are used to evaluate the central difference. In order to determine the off{-}diagonal elements of the stress, the viscosity is to be determined at a cell corner. This is done as follows: At first the cell viscosities of adjacent four cells are calculated according to Eq.~\ref{eq:ViscosityAveraging}. Then viscosities are determined at the cell faces as the arithmetic mean of the neighboring cell values, see left picture of Fig.~\ref{fig:ViscosityAveraging}. Subsequently, corner values are computed using harmonic means of each opposite viscosity values, see right picture of Fig.~\ref{fig:ViscosityAveraging}. The results are two viscosity values at the corner. Finally, the minimum of these two values is chosen as the viscosity at the cell corner. The method of harmonic averaging was introduced by~\cite{Tryggvason1998} in analogy to the calculation of heat conduction.
\section{Results and Discussion}
In contrast to investigations of former authors (\cite{Ash90,Gao05}) who used a frozen image technique with limitations on the dynamics of the collision, the proposed method offers high time resolution and still a very good contrast where no in{-}motion unsharpness occurs. Now it is possible to study not only the dynamic behaviour of the collision complex (outer surface) but also the internal mixing and even the penetration can be analysed quantitatively. The application of a second synchronous high speed camera allows for more detailed insights of the mixing or penetration of the fluid because only the fluorescent material is visible. So, the distribution of the marker is can be determined better, especially for ligaments, penetration or encapsulation of liquid.
\begin{figure}[hbt]
\centering
\includegraphics[width=12 cm]{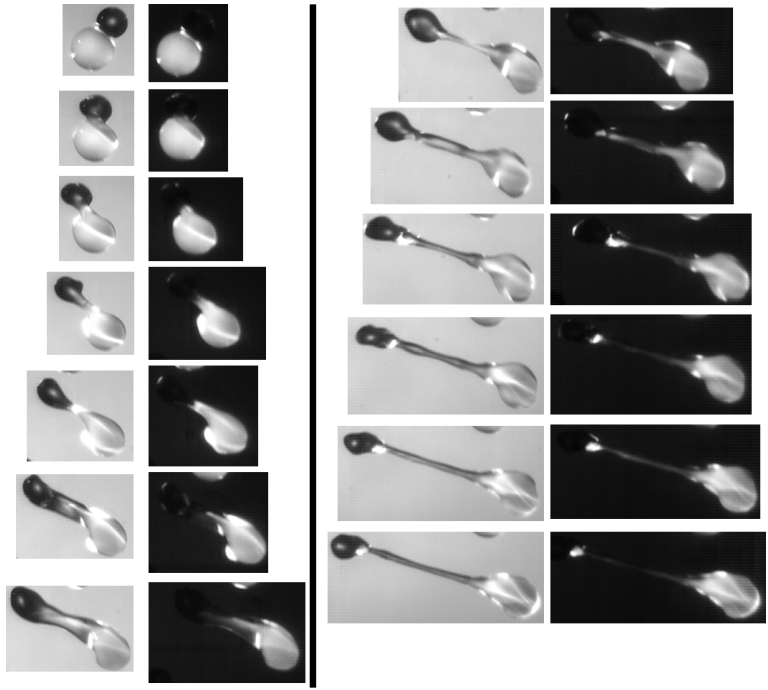}
\caption{Time{-}resolved collision of isoviscous droplets with a $\Delta=0.72$; K30 with 5\% mass fraction for both}
\label{Methode}
\end{figure}

\begin{figure}[hbt]
\centering
\includegraphics[width=10 cm]{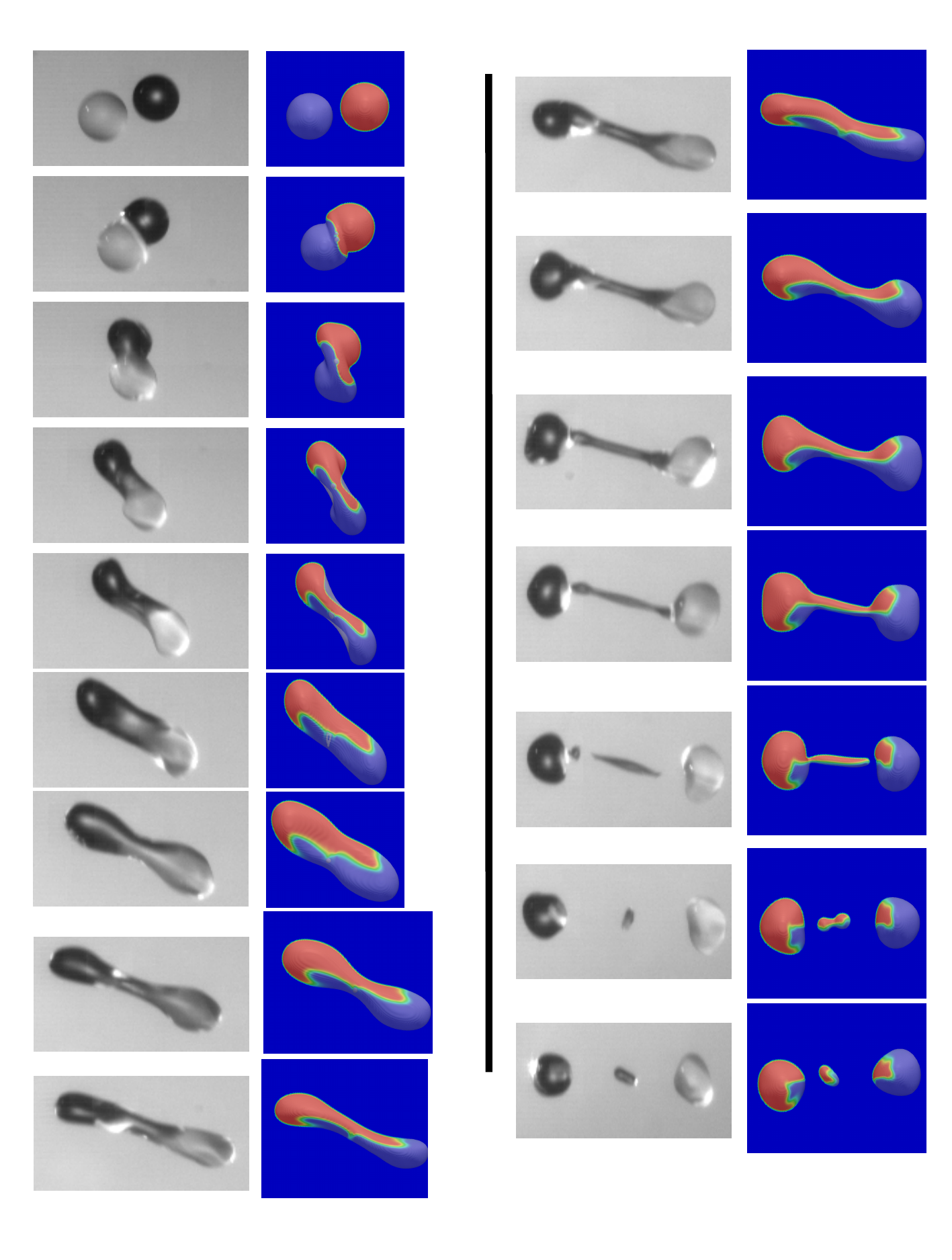} 
\caption{Experimental results (left) and numerical results (right) of a collision of isoviscous droplets.}
\label{fig:IsoviscousExpSim}
\end{figure}
Another advantage which comes with the new method is based on the fact, that numerical researchers now have better validation data due to the high time resolution. 
Fig. \ref{Methode} proves that the developed method works quite well and delivers good results. The interval between two frames is 100 $\mu$s. Both droplets are composed of K30 with 5 \% mass fraction yielding identical viscosity. The transparent droplet carries the marker and glows due to Laser Induced Fluorescence (LIF). It can be seen that at the initial stage a deformation of the droplet occurs and the surface disrupts after 100{-}200 $\mu$s, then a ligament is formed which is stretched until it collapses. Inside the ligament two layers exist, whereas the fluorescent liquid can be found on the bottom of the ligament only. The mixing in this case is quite low, only at the ends of the ligament a blend of both liquids is found.

\begin{figure}[hbt]
\centering
\includegraphics[width=8 cm]{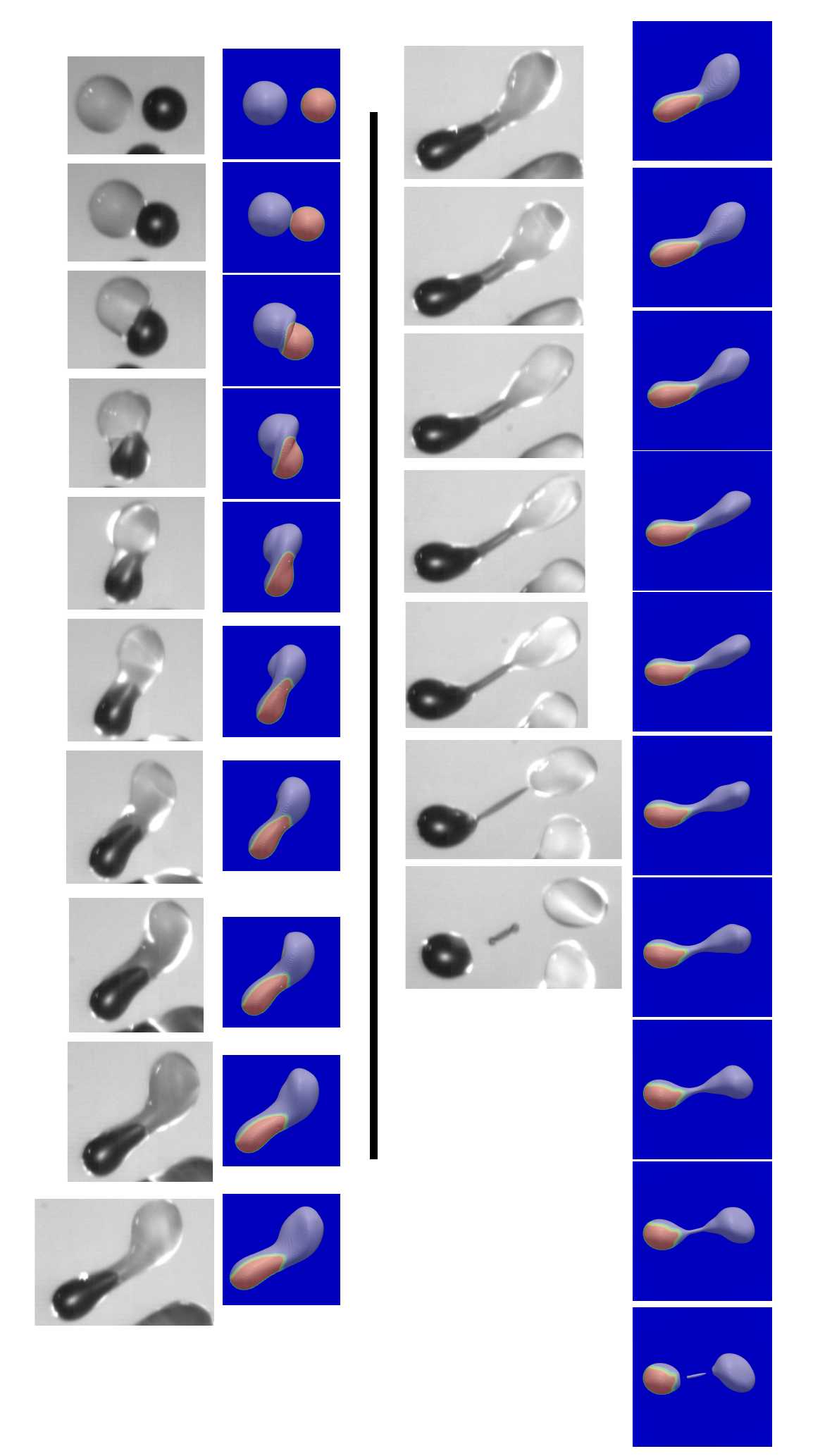}
\caption{Collision of non{-}isoviscous droplets, experimental results left and numerical results right.}
\label{fig:NonIsoviscousExpSim}
\end{figure}
The picture series~\ref{fig:IsoviscousExpSim} shows the collision of isoviscous droplets. The experimental pictures are on the left side and the corresponding numerical results are placed on the right side. The comparison proves that the developed methods work quite well and deliver good results. The interval between two frames is 100 $\mu$s. Both droplets are composed of K30 with 5 \% mass fraction yielding identical viscosity. The transparent droplet carries the marker and glows due to LIF. A mass fraction (red) is transported in the numerical simulation, but it has no influence on the viscosity of the liquid. It can be seen that the two droplets merge directly after the initial contact, then a ligament is formed which is stretched until it collapses. The fluorescent marker can be found in the ligament and in both resulting droplets. The simulation shows good agreement with the experimental data, involving the time{-}scale of the collision as well as the distribution of the mixing fraction.

In contrast to the mixing of two droplets with similar viscosity, a penetration can be observed if the viscosity ratio is raised. The picture series of Fig.~\ref{fig:NonIsoviscousExpSim} shows the collision of droplets with 5\% mass fraction and 25\% mass fraction of K30 respectively; the other parameters are $d \approx 500 \mu m$, $u_{rel}=2.26 m/s $ and $B=0.60$. The left row contains the experimental pictures; the right row displays the visualisation of the numerical results. $4 \cdot 10^6$ cells are used in the simulation, that equals to a resolution of about 64 cells per diameter. The Reynoldsnumber of the high viscous droplet is 435, of the low viscous droplet 135 and of the gas flow 75. In the early stages of the collision the time step size is restricted by the CFL condition while in later stages of the collision the time step size is limited by the viscous momentum diffusion. The frame rate is the same as above, the small droplet (dark) has a much higher viscosity $\frac{\eta_{hv}}{\eta_{lv}}\approx 60$ than the larger fluorescing droplet (bright). The small droplet penetrates into the larger one with time and also large deformations inside the droplet complex occur.

Small differences between the experimental and numerical results can be observed. Comparing the time{-}scale of the collision in experiment and simulation shows that the final rupture of the ligament at the end of the collision is delayed in the simulation, approximately by 0.3 ms. We assume that this delay is attributed to a delay in the coalescence, which is described in more detail below. Experiment and simulation show that the lower viscous liquid tends to encapsulate the higher viscous droplet, as the curvature of the liquid{-}liquid interface indicates. Mixing is not observed, because the contact time is too small compared to the time{-}scale of the diffusion. The result of the collision are two bigger droplets and one small satellite droplet in the middle. The middle droplet and the right satellite droplet are composed only of bright liquid, whereas the left satellite droplet contains a blend of both liquids. Experiment and the numerical simulation show that the droplet on the left contains high viscous liquid which is encapsulated by the low viscous one.

\begin{figure}[hbtp]
\centering
\includegraphics[width=10 cm]{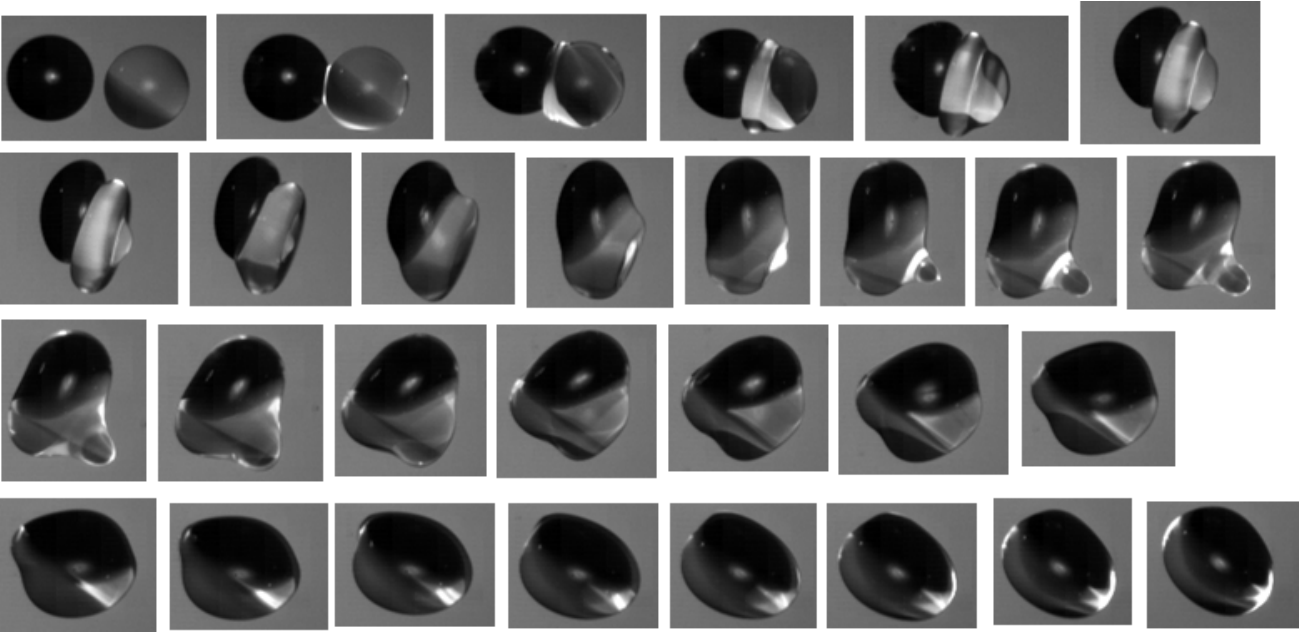}
\caption{Delayed coalescence of different viscous droplets; fluorescent droplet 5\% mass fraction the other droplet with 25\% mass fraction of K30; $d \approx 730 \mu m$, $u_{rel}=1.33 m/s $ and $B=0.135$}
\label{fig:delayedcoalescence}
\end{figure}

\vspace{0 cm}
Furthermore, the effect of different viscous droplets provokes a new very interesting effect. We call this effect $delayed \,coalescence$. This effect is most likely responsible for the deviations from experiment and simulation. It is not jet clear, if an air cushion exists and if more time (or pressure) is needed to disrupt the surface. It occurs at low to moderate relative velocities of the droplets and is somewhat similar to a collision of a droplet with a wall, although not the same. At higher relative velocities the effect is reduced again and plays only a minor role for the collision. The $delayed \,coalescence$ leads to a cusp in the contact region (see the right picture in the upper row of Fig.~\ref{fig:delayedcoalescence}. At a later stage of the collision a partial rebound of the liquid at the opposite collision side of the low viscous droplet appears shortly after the interfaces collapse. Fig.~\ref{fig:delayedcoalescence} shows exemplarily the effect. 

In the numerical method it will not be possible to resolve a thin air gap between the droplets. A possibility to model the effect is to modify the surface tension computation. In the first part of the collision the curvature computation for both droplets will be made separately. At a prescribed time the algorithm is switched to use the complete liquid for the curvature computation and hence the two droplets merge. The modeling of the $delayed \,coalescence$ is part of ongoing work.

\vspace{10 cm}

\section{Summary and Conclusions}
The present work deals with the collision of non{-}isoviscous droplets. The experimental method as well as the numerical code is extended to analyse the collision of different liquids. A new experimental method has been developed in order to visualise the mixing and penetration process of two colliding droplets. Therefore, the fluorescence marker Rhodamine B was feeded to one liquid and droplets were excited by an Ar+ Laser. A combination of LED back light and fluorescence light were recorded with two synchronous cameras, whereas the second camera only recorded the fluorescence in order to have more details on the distribution of the fluorescence marker.
The numerical method is extended to simulate the collision of two non{-}isoviscous droplets by solving an additional transport equation to simulate the mass fraction distribution inside the collision complex. The viscosity is coupled to the mass fraction and a careful averaging is employed at the liquid{-}liquid interface to capture the dynamics of the interface.

The simulation of the collision of isoviscous droplets show that the numerical method is capable to predict the dynamics of the liquid{-}liquid interface inside the collision complex. In case of the non{-}isoviscous droplets, the simulation gives reliable results for the outcome of the collision as well as the distribution of the mass fraction, only small differences in the temporal scale of the collision can be found. 
Two phenomena could be identified which are attributed to the viscosity difference of the droplets. In case of separation, a resulting droplet can occur where the lower viscous droplet encapsulates the higher viscous one. This effect can be observed in the experiments as well as in the numerical data. In case of moderate relative velocities an effect called $delayed \,coalescence$ can be observed. The prediction of this effect by the numerical method is subject of ongoing research.
\section{Acknowledgements}
We gratefully acknowledge financial support provided by the Deutsche Forschungsgemeinschaft within the scope of the Priority Program 1423 ''Process Sprays''.
\section{References}

\end{document}